\documentclass[aps,prd,reprint,groupedaddress,showpacs]{revtex4-1}
\usepackage{amssymb}

\usepackage{graphicx}
\usepackage{dcolumn}
\usepackage{bm}

\begin{document}

\date{\today }
\title{Gravipulsons}
\author{Vladimir A. Koutvitsky}
\author{Eugene M. Maslov}
\email{zheka@izmiran.ru}
\affiliation{Pushkov Institute of Terrestrial Magnetism, Ionosphere and Radio Wave
Propagation of the Russian Academy of Sciences,\\ Troitsk, Moscow
Region, 142190, Russia}
\date{\today}

\begin{abstract}
We search for self-gravitating oscillating field lumps (pulsons) in the
scalar model with logarithmic potential. 
With the use of a Krylov-Bogoliubov-type
asymptotic expansion in the gravitational constant, the pulson solutions of the
Einstein-Klein-Gordon system are obtained in the Schwarzschild coordinates. They
are expressed in terms of solutions of the singular Hill's
equation. The masses of the obtained pulsons are calculated. The initial
conditions are found under which the pulson solutions become periodic. These
conditions are then used in direct numerical integration of the
Einstein-Klein-Gordon system. It is shown that they do evolve into a very
long-lived periodic pulson. Stability of the self-gravitating pulsons and
their possible astrophysical applications are briefly discussed. 
\end{abstract}

\maketitle

\section{Introduction} \label{Sec0}

A large number of modern astrophysical observations suggest the existence of
scalar fields in our Universe as possible candidates for dark matter.
Pulsons are localized configurations of the fields having oscillating energy
density. Numerical simulations of Seidel and Suen \cite%
{{Seidel-Suen1},*{Seidel-Suen2}} have revealed the existence of long-lived
self-gravitating pulsons, so-called oscillating soliton stars or
oscillatons, in the Einstein-Klein-Gordon (EKG) system%
\begin{eqnarray}
R_{\mu \nu }-\frac{1}{2}Rg_{\mu \nu } &=&\varkappa \left[ \phi _{,\mu }\phi
_{,\nu }-\left( \frac{1}{2}\phi _{,\alpha }\phi ^{,\alpha }-U(\phi )\right)
g_{\mu \nu }\right] ,  \nonumber \\
\phi _{;\alpha }^{;\alpha }+U^{\prime }(\phi ) &=&0  \label{eq1}
\end{eqnarray}%
with the potential $U(\phi )=(m^{2}/2)\phi ^{2}$ corresponding to a free
massive scalar field. The authors have established that soliton stars can be
formed from rather general initial field distributions due to specific
relaxation process, the gravitational cooling.

Pulsons were first observed numerically by Bogolubsky and Makhankov \cite%
{{Bog-Makh1},*{Bog-Makh2}} in the Klein-Gordon (KG) model with $\phi ^{4}$
and $sG$ potentials. In these cases, in the absence of gravity, the formation of
the pulsons occurs solely due to self-coupling effects. In the present-day
literature such configurations are often called oscillons, but below we shall use
their original name, pulsons \cite{Bog1}.

Subsequent investigations have shown that pulsons exist in various models and
spatial dimensions, and that they evolve from the diversity of initial conditions \cite%
{Marques, Bog2, Olsen, Geicke, Maslov1, Gleiser1, Copeland, Maslov2, Piette,
Hormuz, Maslov3, Dymnik, Choptuik, Urena, Gleiser2, Kasuya, Koutv1, Koutv2,
Gleiser3} (see \cite{Fodor1}\ for a review). It turns out that pulsons can
arise from both uniform and non-uniform field distributions. Thus pulsons
can emerge in scalar condensates due to the parametric instability of the
spatially uniform background oscillating near a vacuum value \cite{Maslov3,
Gleiser2, Kasuya, Koutv2}. In this case the energy of the background
oscillations is transferred to an incipient pulson via the resonance
mechanism. Quite a different scenario is realized when pulsons are formed
from localized field distributions that appear, e.g., in shrinking
cylindrical domain walls \cite{Geicke}, in collapsing spherical bubbles \cite%
{Gleiser1, Copeland}, or at bubble collisions \cite{Dymnik}. In such a case
an initial field lump sheds excessive energy by radiation of scalar waves
(gravitational cooling of the soliton stars) and settles into a quasi-stable
state, the pulson, whose lifetime depends strongly on the initial
conditions. This suggests the existence of such initial conditions that
evolve into very long-lived quasi-periodic, or even infinitely long-lived
periodic pulsons. The latter would imply\ the existence of exact localized
time-periodic solutions. For the $\phi ^{4}$, $\phi ^{3}-\phi ^{4}$, and $sG$
models, certain of these initial configurations have been found numerically
in \cite{Copeland, Piette, Hormuz, Choptuik}. Recently, in Ref. \cite{Fodor2}
small amplitude pulson solutions of the EKG system have been obtained for
the potentials expansible in a power series. This brings up the following question: 
How does gravity affect the dynamics of the finite amplitude pulsons? For example,
could gravity turn non-periodic pulson solutions into periodic ones?
Consideration of finite amplitude pulsons takes on great significance in the
case where a scalar field potential is not expansible in a power series in
the small amplitude limit.

In this paper we search for pulsons in the{\Large \ }EKG system (\ref{eq1})
with the potential%
\begin{equation}
U(\phi )=\frac{m^{2}}{2}\phi ^{2}\left( 1-\ln \frac{\phi ^{2}}{\sigma ^{2}}%
\right) ,  \label{eq2}
\end{equation}%
where $\phi$ is a real scalar field, $m$ is a bare mass (in units $\hbar =c=1$), and $\sigma $ %
is a characteristic amplitude of the field which is assumed to be finite, but not too
large, so that $\varkappa \sigma ^{2}\ll 1$, where $\varkappa$ is the gravitational constant. 

The nonlinear KG equation with %
the logarithmic potential (\ref{eq2}) was first considered in quantum
field theory by Rosen \cite{Rosen} and later by Bialynicki-Birula and
Mycielski \cite{BBM}. In general, for the nonlinear KG equation the
potential (\ref{eq2}) is the only one which permits real solutions of the
form $\phi =a(t)u(\mathbf{r})$ to exist \cite{Maslov1}. Such singular
potentials currently appear in inflationary cosmology \cite{Barrow} and in
some supersymmetric extensions of the standard model (flat direction
potentials in the gravity mediated supersymmetric breaking scenario) \cite%
{Enqvist}. The logarithmic term in parentheses arises due to quantum
corrections to the bare inflaton mass. 

The paper is organized as follows. In Sec.~\ref{Sec2}, using the smallness of the
gravitational constant, we obtain the approximate solution of the EKG system (\ref{eq1})
which describes time-periodic pulsons of a finite amplitude in the
Schwarzschild metric $ds^{2}=Bdt^{2}-Adr^{2}-r^{2}(d\vartheta ^{2}+\sin
^{2}\vartheta \;d\varphi ^{2})$. In Sec.~\ref{Sec3} we use the obtained solution to find
the initial conditions for direct numerical integration of the system.
We show that these initial conditions do evolve into a very long-lived
periodic pulson. Stability of the self-gravitating pulsons and their possible 
astrophysical meaning are briefly discussed in Sec.~\ref{Sec4}.

\section{Solution}  \label{Sec2}

After the scaling $t\rightarrow t/m,\;r\rightarrow r/m,\;\phi /\sigma
\rightarrow \phi ,$ $\varkappa \sigma ^{2}/2\rightarrow \varkappa ,$ the
system (\ref{eq1}) takes the form%
\begin{equation}
\frac{A_{r}}{A}+\frac{A-1}{r}=\varkappa r\left( \frac{A}{B}\phi
_{t}^{2}+\phi _{r}^{2}+A\phi ^{2}(1-\ln \phi ^{2})\right) ,  \label{eq3}
\end{equation}%
\begin{equation}
\frac{B_{r}}{B}-\frac{A-1}{r}=\varkappa r\left( \frac{A}{B}\phi
_{t}^{2}+\phi _{r}^{2}-A\phi ^{2}(1-\ln \phi ^{2})\right) ,  \label{eq4}
\end{equation}%
\begin{equation}
\frac{A}{B}\phi _{tt}-\phi _{rr}-\frac{2}{r}\phi _{r}+\left( \frac{A}{2B}%
\right) _{t}\phi _{t}+\frac{B}{2A}\left( \frac{A}{B}\right) _{r}\phi
_{r}=A\phi \ln \phi ^{2},  \label{eq5}
\end{equation}%
where $\varkappa \ll 1$ is the rescaled gravitational constant. Looking for
localized solutions, we impose the boundary conditions $\phi (t,\infty
)=0,\;A(t,\infty )=1,\;B(t,\infty )=1,\;\phi _{r}(t,0)=0,\;A(t,0)=1.$

\begin{figure}[htb]
\includegraphics[width=0.35\textwidth]{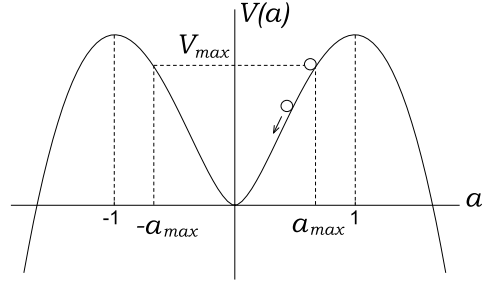}
\caption{The shape of the potential $V(a)$. Initial conditions 
for the nonlinear oscillator (\ref{eq7}) are $a(0)=a_{max},\ a_{t}(0)=0$. }
\label{Fig-1}
\end{figure}

If we set $\varkappa =0$, from (\ref{eq3})-(\ref{eq5}) we immediately obtain%
{\Large \ }$A=B=1$ and arrive at the nonlinear{\Large \ }Klein-Gordon
equation 
\begin{equation}
\phi _{tt}-\phi _{rr}-(2/r)\phi _{r}-\phi \ln \phi ^{2}=0.  \label{eq5a}
\end{equation}
This equation has a whole family of exact pulson solutions \cite{Marques,
Bog2, Maslov1}. The simplest of them is given by%
\begin{equation}
\phi (t,r)=a(t)e^{(3-r^{2})/2},  \label{eq6}
\end{equation}%
where $a(t)$ satisfies the equation of a nonlinear oscillator,%
\begin{equation}
a_{tt}=-dV(a)/da,\quad V(a)=(a^{2}/2)(1-\ln a^{2}).  \label{eq7}
\end{equation}%
As is clear from the shape of the potential $V(a)$ depicted in Fig.~\ref{Fig-1}, oscillations are
possible in the range $-1<a(t)<1$, so we shall consider below that the
pulson's amplitude may be finite, $\left\vert \phi \right\vert \lesssim O(1)$.

For small $\varkappa \ll 1$\ we construct the Krylov-Bogoliubov-type
asymptotic expansion (see, e.g., \cite{Nayfeh}) near the non-gravitating
pulson, 
\begin{eqnarray}
\phi (t,r) &=&\left[ a(\theta )+\varkappa Q(\theta ,r)+O(\varkappa ^{2})%
\right] e^{(3-r^{2})/2},  \label{eq8} \\
\theta _{t} &=&1+\varkappa \Omega +O(\varkappa ^{2}),  \label{eq9}
\end{eqnarray}%
where $a(\theta )$\ satisfies Eqs. (\ref{eq7}), with the phase $\theta $\
instead of $t$, and the initial conditions $a(0)=a_{\max }<1$, $a_{\theta
}(0)=0$. The function $Q(\theta ,r)$ and the constant $\Omega $ to be found
describe the deviation of the pulson's shape from the Gaussian one and the
frequency shift $\delta \omega /\omega =\varkappa \Omega $ due to
gravitational effects.

Setting in Eqs. (\ref{eq3}), (\ref{eq4})%
\begin{equation}
A(t,r)=(1-r_{g}/r)^{-1},\quad B(t,r)=(1-r_{g}/r)e^{-s}  \label{eq9a}
\end{equation}%
and using (\ref{eq8}), (\ref{eq9}), we find%
\begin{eqnarray}
r_{g}(t,r) &=&\varkappa \int_{0}^{r}\left( \frac{1}{B}\phi _{t}^{2}+\frac{1}{%
A}\phi _{r}^{2}+\phi ^{2}(1-\ln \phi ^{2})\right) r^{2}dr  \nonumber \\
&=&\varkappa \left[ V_{\max }\left( (\sqrt{\pi }/2)e^{r^{2}}\mathrm{erf}%
\,r-r\right) -a^{2}r^{3}\right] e^{3-r^{2}}  \nonumber \\
&&+O(\varkappa ^{2}),  \label{eq10} \\
s(t,r) &=&2\varkappa \int_{r}^{\infty }\left( \frac{A}{B}\phi _{t}^{2}+\phi
_{r}^{2}\right) r\,dr  \nonumber \\
&=&\varkappa \left( 2V_{\max }+a^{2}\ln a^{2}+a^{2}r^{2}\right) e^{3-r^{2}} 
\nonumber \\
&&+O(\varkappa ^{2}),  \label{eq11}
\end{eqnarray}%
where $a=a(\theta (t))$, $V_{\max }=V(a_{\max })$. Substituting (\ref{eq8})
into (\ref{eq5}) leads to the equation for $Q(\theta ,r)$:%
\begin{equation}
Q_{\theta \theta }-Q_{rr}+(2/r)(r^{2}-1)Q_{r}-(2+\ln a^{2})Q=S(a,r),
\label{eq11a}
\end{equation}%
where%
\begin{eqnarray}
S(a,r) &=&a\{V_{\max }[\sqrt{\pi }(2-r^{2}-\ln a^{2})(2r)^{-1}e^{r^{2}}%
\mathrm{erf}\,r \nonumber \\
&&+\,3r^{2}-4-3\ln a^{2}] -5a^{2}r^{2}+2a^{2}r^{4} \nonumber \\
&&+\,a^{2}-2a^{2}\ln ^{2}a^{2}\}e^{3-r^{2}}-2\Omega
a\ln a^{2}.  \label{eq11b}
\end{eqnarray}%
Its solution is given by%
\begin{equation}
Q(\theta ,r)=\frac{1}{r}\sum_{n=0}^{\infty }c_{n}X_{n}(\theta )H_{2n+1}(r),
\label{eq12}
\end{equation}%
where $c_{n}=\pi ^{-1/4}\left[ 2^{2n+1}\left( 2n+1\right) !\right] ^{-1/2}$, 
and $H_{2n+1}(r)$ are Hermite polynomials. The functions $X_{n}(\theta )$ must
satisfy the non-homogeneous singular Hill's equation%
\begin{equation}
X_{n_{\,\scriptstyle{\theta \theta }}}+\left( E-2-\ln a^{2}\right)
X_{n}=f_{n}(a),  \label{eq13}
\end{equation}%
where $E=E_{n}=4n$,%
\begin{equation}
f_{n}(a)=2c_{n}\int_{0}^{\infty}S(a,r)H_{2n+1}(r)e^{-r^{2}}r\,dr.  \label{eq13a}
\end{equation}%
The calculation gives%
\begin{eqnarray}
f_{0}(a) &=&D_{0}(a)-\sqrt{2}\pi ^{1/4}\Omega a\ln a^{2},  \label{eq14} \\
f_{n}(a) &=&D_{n}(a)\quad (n=1,2,...),  \label{eq15} \\
D_{n}(a) &=&\frac{(-1)^{n}(2n)!c_{n}}{2^{n+4}(2n-1)n!}\sqrt{\frac{\pi }{2}}%
e^{3}  \nonumber \\
&&\times \{a^{3}(4n^{2}-1)(4n^{2}-4n-7-16\ln ^{2}a^{2})  \nonumber \\
&&-2V_{\max }a[24n^{3}+20n^{2}-46n-1  \nonumber \\
&&+4(2n-1)(6n+5)\ln a^{2}]\}.  \label{eq16}
\end{eqnarray}%
Note that $f_{n}(a)$ is a $T$-periodic function of $\theta $, while $\ln
a^{2}$ on the left hand side of Eq. (\ref{eq13}) is a $T/2$-periodic one, where $T$
is a period of $a(\theta )$.

Solutions of the homogeneous singular Hill's equation were investigated in
Ref. \cite{Koutv2}. In accordance with the Floquet theory (see, e.g., \cite{Whitt}) 
Eq. (\ref{eq13}) with $f_{n}=0$ has two linearly
independent solutions of the form $\varphi (\theta )e^{\mu \theta }$ and $%
\varphi (-\theta )e^{-\mu \theta }$, where $\mu $ is a characteristic
exponent, and $\varphi (\theta )$ is a $T$-periodic ($T/2$-periodic or $T/2$%
-antiperiodic) function. Obviously, we can set $\varphi (0)=1$. Let $X^{\pm
}\left( \theta \right) $ be two solutions of the homogenious Eq. (\ref{eq13}) (with $f_{n}=0$)
satisfying the conditions $X^{+}\left( 0\right) =1$, $X_{_{\,\scriptstyle{%
\theta }}}^{+}\left( 0\right) =0$, $X^{-}\left( 0\right) =0$, $X_{_{\,%
\scriptstyle{\theta }}}^{-}\left( 0\right) =1$. They can be written as%
\begin{eqnarray}
X^{+}\left( \theta \right) &=&\frac{1}{2}\left[ \varphi (\theta )e^{\mu
\theta }+\varphi (-\theta )e^{-\mu \theta }\right] ,  \label{eq16a} \\
X^{-}\left( \theta \right) &=&\frac{1}{2\left( \mu +\varphi _{_{\,%
\scriptstyle{\theta }}}(0)\right) }\left[ \varphi (\theta )e^{\mu \theta
}-\varphi (-\theta )e^{-\mu \theta }\right] .  \label{eq16b}
\end{eqnarray}%
If $\left| X^{+}\left( T/2\right) \right| >1$, we have the resonance case: $%
\mu >0$ and is determined by the equation $\cosh (\mu T/2)=\left|
X^{+}\left( T/2\right) \right| $, $\varphi (\theta )$ is a real $T/2$%
-periodic or $T/2$-antiperiodic function, and hence oscillations of $X^{\pm
}\left( \theta \right) $ grow exponentially with $\theta $. If $\left|
X^{+}\left( T/2\right) \right| <1$, we have the non-resonance case: $\mu
=i\nu $, and $\varphi (\theta )$ is a complex $T/2$-periodic function such that $%
\varphi ^{\ast }(\theta )=\varphi (-\theta )$. Hence the solutions $X^{\pm
}\left( \theta \right) $ are bounded. They can be periodic (with some
period), or non-periodic depending on $\nu $, which is determined by $\cos (\nu
T/2)=X^{+}\left( T/2\right) $. These cases are realized in different domains
of the $(E,a_{\max }^{2})$ plane that make up a stability-instability chart.
The domains with $\mu >0$ are known as resonance zones. The special case $%
\left| X^{+}\left( T/2\right) \right| =1$ is realized on their boundaries
where $\mu =0$. Then one of the solutions, either $X^{+}\left( \theta \right) $ or 
$X^{-}\left( \theta \right) $,
is a $T$-periodic ($T/2$-periodic or $T/2$-antiperiodic) function, and
another one is proportional to the product of this function times $\theta $
plus some $T$-periodic function ($T/2$-periodic or $T/2$-antiperiodic,
respectively).

The surface $\mu (E,a_{\max }^{2})$ over the resonance zones has been
constructed in Ref. \cite{Koutv2}. For discrete $E=4n$ the above functions
acquire the subscript $n$, so we shall write $\varphi _{n}(\theta )$, $%
X_{n}^{\pm }\left( \theta \right) $, $\mu _{n}(a_{\max }^{2})$. 

\begin{figure}[htb]
\includegraphics[width=0.4\textwidth]{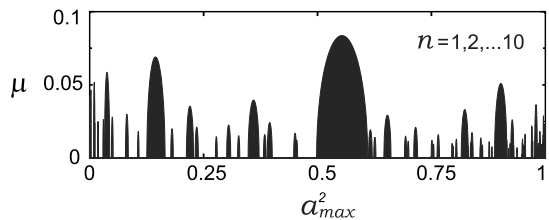}
\caption{A collection of the resonance peaks obtained by superposition of the functions $\mu_n(a_{max}^ 2)$.}
\label{Fig0}
\end{figure}

Each cross-section of the surface $\mu (E,a_{\max }^{2})$ with the plane $E=4n,\ n=1,2,...$, 
gives the characteristic exponent $\mu_{n}$  as a function of $a_{\max }^{2}$.
This function is represented by a series of peaks $\mu_{n}>0$ separated 
by intervals of stability. 
By superposing the curves $\mu_{n}(a_{\max }^{2})$ 
for all considered modes $n=1,2,...,N$, one gets the pattern shown in Fig.~\ref{Fig0}.
The mode $n=0$ corresponds to the above special case $\mu=0$ and thus does not contribute 
to the pattern. 
The obtained composite plot gives an idea of the existence of unstable and (quasi)stable modes 
in different regions of the $a_{\max }^{2}$ axis and demonstrates the tendency 
to progressively fill the interval $0<a_{\max }^{2}<1$ by the resonant peaks 
as the successively higher energy levels $E_{n}=4n$ are accounted for.

In terms of $X_{n}^{\pm }\left( \theta \right) $ the general solution of Eq.
(\ref{eq13}) is written as%
\begin{eqnarray}
X_{n}\left( \theta \right)  &=&\left( X_{n}\left( 0\right) -\int_{0}^{\theta
}X_{n}^{-}f_{n}\,d\theta \right) X_{n}^{+}\left( \theta \right)   \nonumber
\\
&&\hspace{-10pt}+\left( X_{n_{\,\scriptstyle{\theta }}}\left( 0\right)
+\int_{0}^{\theta }X_{n}^{+}f_{n}\,d\theta \right) X_{n}^{-}\left( \theta
\right) .  \label{eq17}
\end{eqnarray}%
The solutions $X_{0}^{\pm }\left( \theta \right) $\ have the form%
\begin{eqnarray}
X_{0}^{+}\left( \theta \right)  &=&\xi q^{-1/3}(\xi ^{2})-\xi _{\theta
}\int_{0}^{\theta }K(\xi ^{2})\,d\theta ,  \label{eq18} \\
X_{0}^{-}\left( \theta \right)  &=&-(\omega _{0}^{2}-1)^{-1}\xi _{\theta },
\label{eq19}
\end{eqnarray}%
where$\;$the notations $\xi (\theta )=a/a_{\max }$, $\omega
_{0}^{2}=1-\ln a_{\max }^{2}$ are introduced,%
\begin{equation}
\xi _{\theta }^{2}=(\omega _{0}^{2}-1)(1-\xi ^{2})q^{-2/3}(\xi ^{2})
\label{eq19a}
\end{equation}%
is the first integral of Eq. (\ref{eq7}) in terms of $\xi (\theta )$%
, and the functions $q(\xi ^{2})$\ and $K(\xi ^{2})$\ are%
\begin{eqnarray}
q(\xi ^{2}) &=&\left( \frac{\omega _{0}^{2}-1}{\omega _{0}^{2}+(1-\xi
^{2})^{-1}\xi ^{2}\ln \xi ^{2}}\right) ^{3/2},  \label{eq20} \\
K(\xi ^{2}) &=&\frac{1-q(\xi ^{2})}{1-\xi ^{2}}q^{-1/3}(\xi ^{2}).
\label{eq21}
\end{eqnarray}%
Note that $0<\left( 1-\omega _{0}^{-2}\right) ^{3/2}\leqslant q(\xi
^{2})\leqslant 1$, $dq/d\xi ^{2}>0\;(\xi ^{2}\leqslant 1)$, and $q(1)=1$. 
Since $K(\xi ^{2})$ is a sign-definite periodic
function of $\theta $, its average $\overline{K}%
=T^{-1}\int_{0}^{T}K\,d\theta \neq 0$, so the solution (\ref%
{eq18}) can be represented in the form $X_{0}^{+}\left( \theta \right) =-%
\overline{K}\xi _{\theta }\theta +\psi (\theta )$, where $\psi
(\theta )$ is a $T/2$-antiperiodic function with $%
\overline{\psi (\theta )}=0$ [here and elsewhere the bar means
the average over the period $T$ of $a(\theta )$]. Thus
oscillations of $X_{0}^{+}\left( \theta \right) $ grow linearly
with $\theta $ for any $a_{\max }$. This is in agreement
with the fact that $X_{0}^{+}\left( T/2\right) =-1$ and the line $%
E=0$ is the boundary of a resonance zone on the $(E,a_{\max }^{2})$ %
plane \cite{Koutv2}. The equality $X_{0}^{+}\left( T/2\right) =-1$ %
immediately follows from Eq. (\ref{eq18}) if one takes into account that $%
\xi (T/2)=-1$, $\xi _{\theta }(T/2)=0$ (see Fig.~\ref{Fig-1}).

The requirement of boundedness of the general solution $X_{0}\left(
\theta \right)\ $(\ref{eq17}) determines $\Omega $ and,
hence, the frequency shift in accordance with Eq. (\ref{eq9}). Indeed,
substituting $X_{0}^{+}\left( \theta \right) $, $X_{0}^{-}\left(
\theta \right)$, and $f_{0}(a)$ into Eq. (\ref{eq17}) and
integrating by parts, we find that the linearly growing terms cancel out if%
\begin{equation}
\Omega =\frac{\sqrt{2}}{\pi ^{1/4}a_{\max }}\left( \frac{\overline{%
X_{0}^{+}\left( \theta \right) D_{0}(a)}}{\ln a_{\max }^{2}}-X_{0}\left(
0\right) \overline{K}\right).  \label{eq22}
\end{equation}%
 Under the condition (\ref{eq22}%
) the solution $X_{0}\left( \theta \right) $ is a bounded $T$-periodic
function.

To obtain the corresponding conditions for $n\geqslant 1$, we substitute (\ref%
{eq16a}), (\ref{eq16b}) into (\ref{eq17}) and require that $X_{n}\left(
\theta \right) =X_{n}\left( \theta +T\right) $. In this equality the
integrals between the limits $0$ and $\theta $ cancel
out. The remaining terms make up a linear combination of the independent
solutions $\varphi _{n}(\theta )e^{\mu _{n}\theta }$ and $\varphi
_{n}(-\theta )e^{-\mu _{n}\theta }$. Equating to zero coefficients
of these solutions and using the identities%
\begin{eqnarray}
e^{-\mu _{n}T/2}\overline{\varphi _{n}(\theta )e^{\mu _{n}\theta }f_{n}(a)}
&=&e^{\mu _{n}T/2}\overline{\varphi _{n}(-\theta )e^{-\mu _{n}\theta
}f_{n}(a)}  \nonumber \\
&=&\frac{\overline{X_{n}^{+}\left( \theta \right) f_{n}(a)}}{\cosh \left(
\mu _{n}T/2\right) },  \label{eq22a}
\end{eqnarray}%
\begin{equation}
\mu _{n}+\varphi _{n_{\,\scriptstyle{\theta }}}(0)=\frac{X_{n_{\,\scriptstyle%
{\theta }}}^{+}\left( T\right) }{\sinh \left( \mu _{n}T\right) },
\label{eq22b}
\end{equation}%
we arrive at the conditions%
\begin{eqnarray}
X_{n}\left( 0\right) &=&-\frac{T}{X_{n_{\,\scriptstyle{\theta }}}^{+}\left(
T\right) }\overline{X_{n}^{+}\left( \theta \right) f_{n}(a)},  \label{eq23}
\\
X_{n_{\,\scriptstyle{\theta }}}\left( 0\right) &=&0.  \label{eq23a}
\end{eqnarray}%
Note that $X_{n_{\,\scriptstyle{\theta }}}^{+}\left( T\right) \neq 0$
because $\mu _{n}+\varphi _{n_{\,\scriptstyle{\theta }}}(0)$ in (\ref{eq22b}) 
is proportional to the Wronskian $W\left( \varphi _{n}(\theta )e^{\mu
_{n}\theta },\ \varphi _{n}(-\theta )e^{-\mu _{n}\theta }\right) $. Equation (\ref%
{eq22b}) can be easily derived if one expresses  $\varphi _{n}(\theta )$ from (%
\ref{eq16a}) in terms of $X_{n}^{+}\left( \theta \right) $ and $%
X_{n}^{+}\left( \theta +T\right) $ and takes into account that $%
X_{n}^{+}\left( T\right) =\cosh \left( \mu _{n}T\right) $.

Interestingly, Eq. (\ref{eq23}) is still valid on the boundaries of
resonance zones, Eq. (\ref{eq23a}) being no longer necessary. In particular,
this is true for $n=0$. Indeed, differentiation of (\ref{eq18}) gives $%
X_{0_{\,\scriptstyle{\theta }}}^{+}\left( T\right) =-T\overline{K}\ln
a_{\max }^{2}$. To calculate $\overline{X_{0}^{+}\left( \theta \right)
f_{0}(a)}$ we substitute $a\ln a^{2}=a_{\theta \theta
}=a_{\max }\ln a_{\max }^{2}X_{0_{\,\scriptstyle{\theta }}}^{-}\left( \theta
\right) $ in (\ref{eq14}) and, integrating by parts, take into account that $W\left(
X_{0}^{+}\left( \theta \right),\ X_{0}^{-}\left( \theta \right) \right) =1$.
As a result, we arrive at the condition (\ref{eq22}) again.

Thus, under the conditions (\ref{eq22}), (\ref{eq23}), (\ref{eq23a}) the
solution (\ref{eq12}) is $T$-periodic with respect to $\theta $. This means
the solution (\ref{eq8}) is also periodic [with the period $%
(1+\varkappa \Omega )^{-1}T$ with respect to $t$]. Note it involves the free
parameters $a_{\max }$, $X_{0}\left( 0\right) $, and $X_{0\,_{\scriptstyle{%
\theta }}}(0)$.

To be certain that the obtained solution is correct, we examine the mass
conservation law. The mass of a self-gravitating field lump is defined as $%
M=4\pi \int_{0}^{\infty }T_{0}^{0}r^{2}dr$, where $T_{0}^{0}$ is the energy
density of the scalar field involved in the EKG system (\ref{eq1}). In terms
of the rescaled variables it can be written as $M=(2\pi \sigma
^{2}/m)\lim_{r\rightarrow \infty }(r_{g}(t,r)/\varkappa )$, where $r_{g}(t,r)
$ is defined in (\ref{eq10}), $\varkappa $ being the rescaled gravitational
constant. This limit must be time independent. To check this, we substitute
the solution (\ref{eq8}) into (\ref{eq10}) and calculate the limit of $%
r_{g}/\varkappa $ in the first order in $\varkappa $ using the orthogonality of
the Hermite polynomials. The result is given by%
\begin{eqnarray}
M&=&\frac{\left( e\sqrt{\pi }\right) ^{3}\sigma ^{2}V_{\max }}{m}\Biggl\{ 1-\varkappa \frac{\sqrt{2}a_{\max }}{\pi ^{1/4}V_{\max }}\biggl[ X_{0}\left(
0\right) \ln a_{\max }^{2} \nonumber \\
&-&\frac{e^{3}\pi ^{1/4}}{128}a_{\max }^{3}\left(1+14\ln a_{\max }^{2}\right) \biggr]+O(\varkappa ^2)\ \Biggr\} ,  \label{eq23b}
\end{eqnarray}%
which is evidently constant.

Since $\varkappa \ll 1$, the gravitational field created by this mass is weak,
as is clearly seen from (\ref{eq9a})-(\ref{eq11}). In the limit $\varkappa
\rightarrow 0$\ the gravity vanishes. However, the rescaled scalar field
persists, satisfying Eq. (\ref{eq5a}), and its amplitude may have any value
in the range $0<a_{\max }<1$. As $a_{\max }$ changes from unity to zero, the
pulson's frequency changes from zero to infinity, correspondingly. In
particular, in the small amplitude limit the pulson's frequency is $\omega
_{0}=\left( 1-\ln a_{\max }^{2}\right) ^{1/2}$. 

Thus, we have obtained a three-parametric family of the spatially localized
time-periodic solutions (\ref{eq8})-(\ref{eq11}) of the system (\ref{eq3})-(%
\ref{eq5}), wherein only the smallness of the rescaled gravitational constant $%
\varkappa \;$has been. Note that the smallness of the pulson's
amplitude, $a_{\max }\ll 1,$\ is not assumed in the above consideration. To
our knowledge, this is the first example of the pulson solutions of the EKG
system, that have an arbitrary frequency. We have named them
gravipulsons.\bigskip 

\section{Numerical simulation}  \label{Sec3}

Our solution, however, is an approximate one. It was obtained in the first
order in the gravitational constant. Hence its deviation from an exact solution
increases in time, as happens with any asymptotic solution in the theory of
nonlinear oscillations \cite{Nayfeh}. But we can go back in time and take
the initial state of the obtained solution as initial conditions for direct
numerical integration of the starting EKG system. As a result, we have a
three-parametric family of the initial conditions:%
\begin{eqnarray}
\phi (0,r) &=&a_{\max }e^{(3-r^{2})/2}+\varkappa G(r;a_{\max },X_{0}\left(
0\right) ),   \label{eq24a} \\ 
\phi _{t}(0,r) &=&2\varkappa c_{0}X_{0\,_{\scriptstyle{\theta }}}(0)e^{(3-r^{2})/2}.  \label{eq24}
\end{eqnarray}%
The function $G(r;a_{\max },X_{0}\left( 0\right) )=Q(0,r)e^{(3-r^{2})/2}$
describes admissible deformations of the initial pulson's profile which
evolve into periodic solutions. In calculating $G$ we assume that $a_{\max }$
belongs to one of the intervals of quasi-stability \cite{Koutv1, Koutv2}
where $X_{n}^{\pm }\left( \theta \right) $, with $1\leqslant n\leqslant N$, 
are bounded for sufficiently large $N$. This can be easily inspected by
numerical integration of the Hill's equation, taking into account that
the boundedness of $X_{n}^{\pm }\left( \theta \right) $ is equivalent to the
condition $\left| X_{n}^{+}\left( T/2\right) \right| <1$. We restrict
ourselves to the summation from $0$ to $N$ in (\ref{eq12}). In deciding on $%
N $, it is necessary to take into account that the related error in $Q$ must
not exceed $O(\varkappa )$. Below we take $\varkappa =0.005$, $N=9$, and set 
$X_{0}\left( 0\right) =X_{0\,_{\scriptstyle{\theta }}}(0)=0$ for simplicity.

\begin{figure}[htb]
\includegraphics[width=0.45\textwidth]{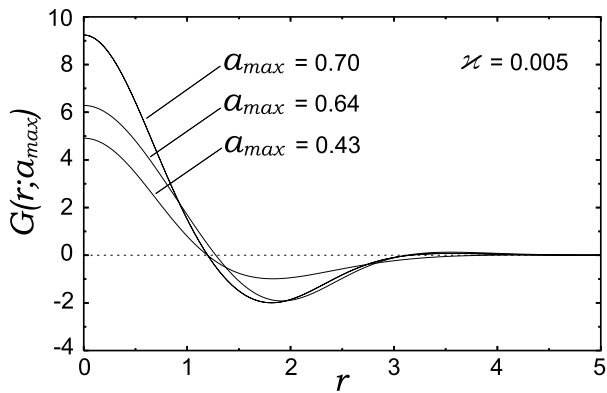}
\caption{Admissible deformations of the initial pulson's profile calculated
by formula (\ref{eq12}) with $X_{0}(0)=0$ and $X_{n}(0)$ (\ref{eq23}).}
\label{Fig1}
\end{figure}

Figure \ref{Fig1} shows the examples of the admissible deformations calculated
for three different values of $a_{\max }$.

\begin{figure*}[thb]
\includegraphics[width=0.9\textwidth]{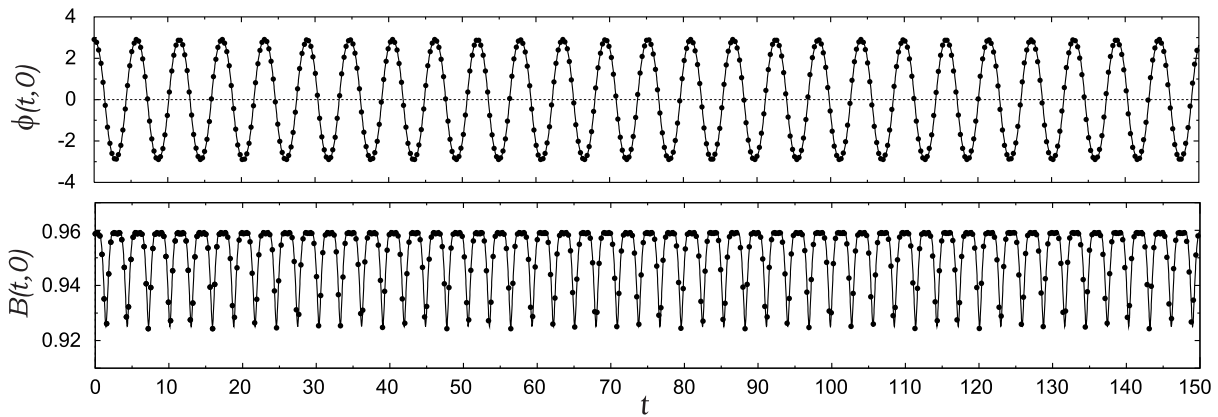}
\caption{Oscillations of the scalar field (top panel) and metric coefficient
(bottom panel) at the center of the pulson for $\varkappa = 0.005,\\
a_{max}=0.64$.} 
\label{Fig2}
\bigskip
\end{figure*}

\begin{figure}[htb]
\includegraphics[width=0.45\textwidth]{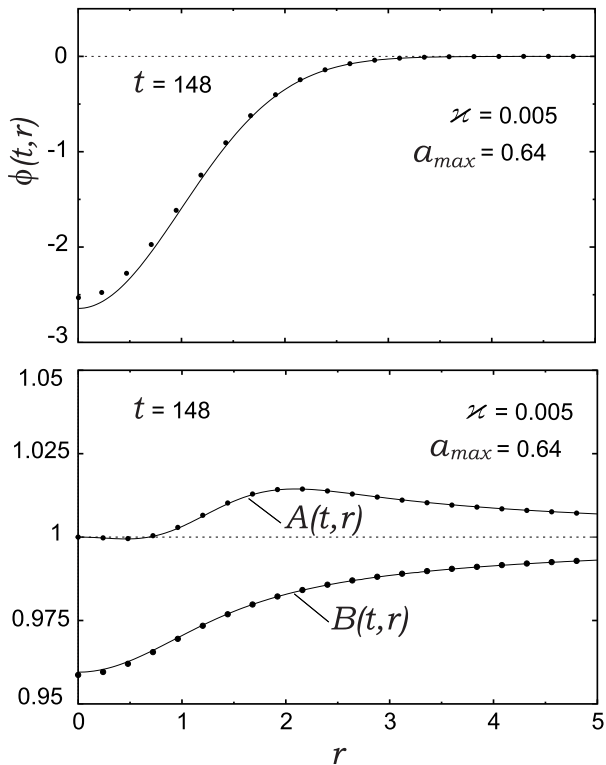}
\caption{Profiles of the scalar field and metric coefficients.}
\label{Fig3}
\end{figure}

In Figs. \ref{Fig2} and \ref{Fig3} we compare our solution (\ref{eq8})-(%
\ref{eq11}) (solid lines) with the results of direct numerical integration
of the EKG system (indicated by dots). We started with one of the
admissible deformations of the pulson's profile that we have found (see Fig. \ref%
{Fig1}). 
Oscillations of the scalar field and metric at the center of the pulson are
shown in Fig.~{\ \ref{Fig2}}. Figure \ref{Fig3} shows the pulson's and metric's 
profiles taken in some intermediate moment of time. 

\begin{figure}[tbh]
\includegraphics[width=0.425\textwidth]{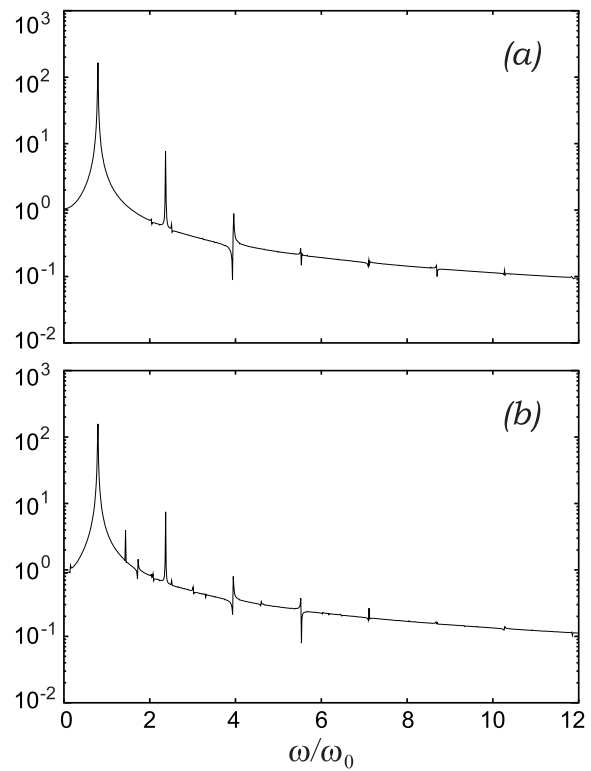}
\caption{Fourier spectrum of $\protect\phi (t,0)$ for admissible~(a) and
inadmissible~(b) initial conditions.}
\label{Fig4}
\end{figure}

We have performed the Fourier analysis of the
scalar field oscillations obtained by numerical integration of the EKG
system. The resulting spectrum shown in Fig. \ref{Fig4}(a) demonstrates
periodicity with high accuracy.

Then we violated the condition (\ref{eq23}) by tripling $X_{1}\left(0\right) $ 
that was calculated before, and integrated the EKG system again. As expected, 
the resulting field oscillations were found to be non-periodic.
The corresponding spectrum is presented in Fig. \ref{Fig4}(b).
Nonperiodicity manifests itself as additional peaks in the spectrum which
are absent in Fig. \ref{Fig4}(a).

\begin{figure}[tbh]
\includegraphics[width=0.45\textwidth]{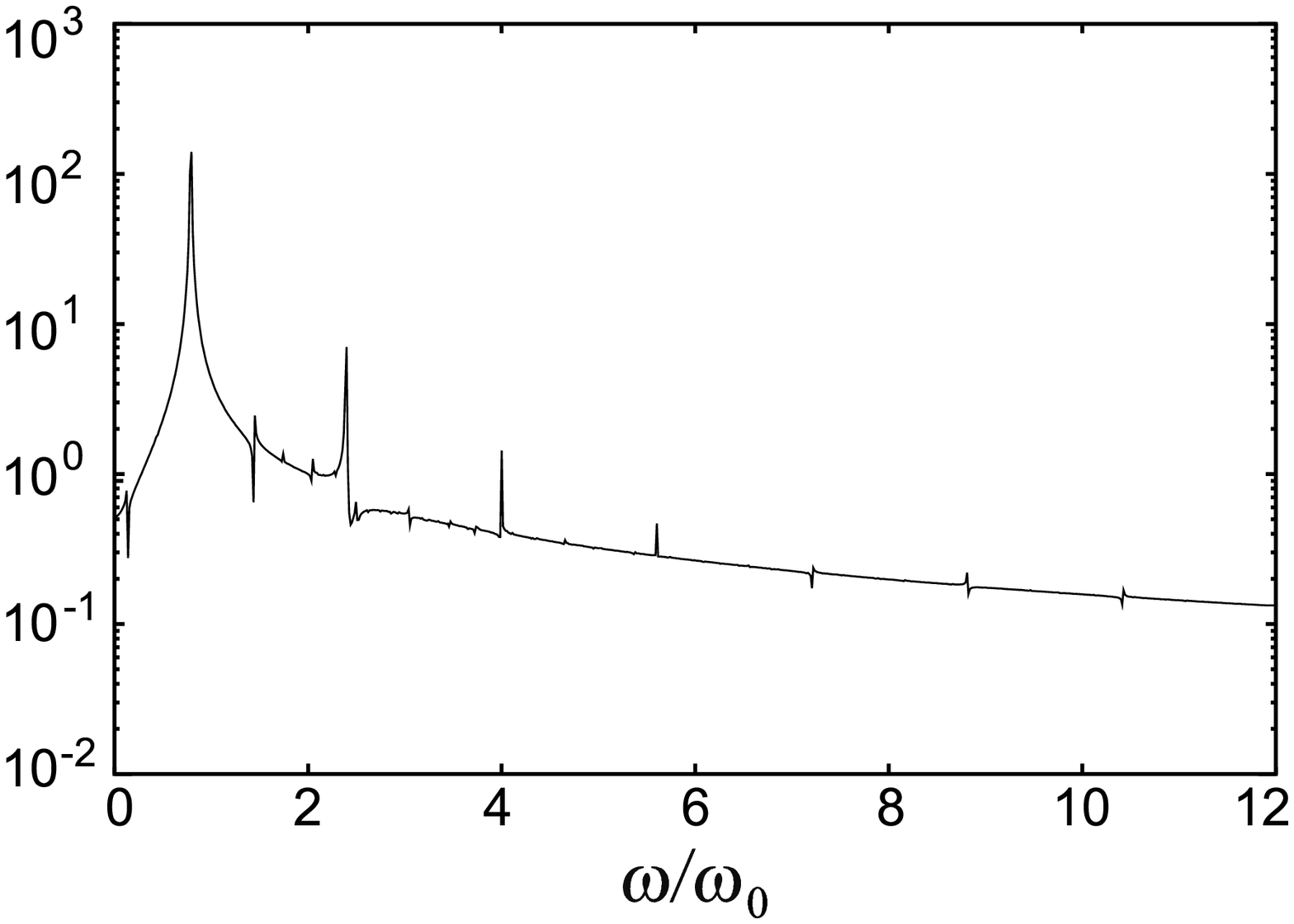}
\caption{Fourier spectrum of $\protect\phi (t,0)$ obtained from the solution 
of the nonlinear KG equation (\ref{eq5a}) with the initial conditions (\ref{eq24a}), 
(\ref{eq24}). All parameters are the same as in Figs. (\ref{Fig2}), (\ref{Fig3}). }
\label{Fig5}
\end{figure}

To clarify the meaning of gravity, we used the obtained initial conditions (\ref{eq24a}), (\ref{eq24}) 
with $\varkappa $ as a formal parameter for the numerical
integration of the nonlinear KG equation (\ref{eq5a}). The solution was
found to be non-periodic, as is clear from its spectrum which is shown in Fig. \ref%
{Fig5}. We thus conclude that it is because of gravity that the periodic
pulsons of the considered non-Gaussian shapes exist.

\section{Concluding remarks}  \label{Sec4}

Thus we have demonstrated the existence of long-lived time-periodic pulsons
in the EKG system. These pulsons differ from the non-gravitating ones in
their shapes and frequencies and exist only due to gravitational effects.

The question arises as to whether these gravipulsons are stable. While the
stability analysis is out of the scope of the present work, it is worth noting that
the stability of the solution (\ref{eq8}), and hence (\ref{eq9a}), is determined by the stability of the
solutions of the non-homogeneous Hill's equation $X_{n}(\theta )$ involved
in (\ref{eq12}). In turn, as it follows from (\ref{eq17}), the stability of the
general solution $X_{n}(\theta )$ is determined by the behavior of the
functions $X_{n}^{+}(\theta )$ and $X_{n}^{-}(\theta )$.

It is clear that all solutions $X_{n}(\theta )$ satisfying the initial
conditions (\ref{eq23}), (\ref{eq23a}) are unstable in the resonance case $%
\mu _{n}>0$. Indeed, any perturbation of the initial values $X_{n}(0)$, $%
X_{n\,_{\scriptstyle{\theta }}}(0)$, determined by (\ref{eq23}), (\ref{eq23a}%
), leads to the appearance of terms $\sim \exp (\mu _{n}\theta )$ on the
right-hand side of (\ref{eq17}), thus making the corresponding function $%
X_{n}(\theta )$, and hence the solution (\ref{eq8}), exponentially growing in
time.

On the other hand, in the non-resonance case $\mu _{n}=i\nu _{n}$, the functions $%
X_{n}^{+}(\theta )$, $X_{n}^{-}(\theta )$ as well as $X_{n}(\theta )$ in (%
\ref{eq17}) are bounded, and a small perturbation of the initial conditions (%
\ref{eq23}), (\ref{eq23a}) results in the appearance of only small oscillating
terms in $X_{n}(\theta )$. So we can expect that the solution (\ref{eq8}) is
stable if all modes $X_{n}(\theta )$ are non-resonant. The question is, does
any value of $a_{\max }$ exist such that all modes $E_{n}=4n\ (n>0)$
are stable?

A collection of the peaks $\mu _{n}>0$ with $n=1,2,...,10$, shown in Fig. %
\ref{Fig0}, demonstrates the existence of numerous stability intervals
separating the instability ones on the $a_{\max }^{2}$ axis. All modes %
$n=1,2,...,10$ are stable in the gaps between the peaks, and thus $\mu_{n}=i\nu _{n}$.  
However, if we take into consideration additional modes
with $10<n\leqslant N$, supplementary peaks must be added to this plot. Some of the
new peaks will be overlapped by the existing ones, but the rest will fall
within the stability intervals and erode them. Nevertheless, narrow
stability gaps remain visible on the abscise axis even in the case of large $%
N$.

While we have no proof that some gaps of stability survive as $N$ goes to
infinity, one should take into account that the amplitude of the peaks in Fig. %
\ref{Fig0} decreases with increasing $n$, and in any case, narrow intervals
on the $a_{\max }^{2}$ axis can be found where only high-$n$ modes are
unstable. We refer to them as intervals of quasi-stability. Indeed, while 
the solution (\ref{eq8}) with $a_{\max }$ falling in one of
these intervals is unstable, this instability evolves very slowly, and
the gravipulson still remains a long-lived object.
Moreover, as it was demonstrated in our simulation \cite{Koutv1}, in the
case of the non-gravitating pulson, the nonlinear stage of instability
saturates very quickly, resulting in a slightly modified pulson which remains
a compact oscillating object. We expect the same instability behavior
in the case of gravipulsons also, at least at small $\varkappa$, even if this
instability is caused by the action of some other perturbative objects around them.

A few words about possible astrophysical applications of the obtained solution are in order. 
There are a number of papers where scalar solitons are considered
as models of galactic halos in hopes of explaining the observational
flatness of the rotation curves (see., e.g., \cite{Mielke} and references therein). %
It is easy to see that then the
energy density of a scalar field must not decay faster than $r^{-2}$.
Evidently, our solution does not satisfy this criterion. However, if a
galactic halo is not a single soliton-like object, but is an ensemble 
of dark matter lumps, of so-called ''mini-MACHOs'' \cite{Hernandez}%
, the gravipulsons may be reasonable candidates for these compact
constituents. In this case the gravipulson masses (\ref{eq23b}) need to be
limited by the condition $M\lesssim 10^{-7}M_{\odot }$ following from
microlensing data \cite{Alcock}. This constrains the amplitude of the
gravipulsons and the parameters of the potential (\ref{eq2}).

\begin{acknowledgments}
We are grateful for discussions with participants of the IV International Conference 
"Frontiers of Nonlinear Physics" (FNP-2010). 
\end{acknowledgments}

\nocite{*}
\bibliography{Gravipulson}
\bigskip

\end{document}